\documentclass[journal]{IEEEtran}
\usepackage{blindtext}
\usepackage{graphicx}

\usepackage{cite}
\ifCLASSINFOpdf
\else
\fi
\usepackage{multirow}

\usepackage[cmex10]{amsmath}
\usepackage{array}
\usepackage[tight,footnotesize]{subfigure}
\usepackage[caption=false,font=footnotesize]{subfig}
\begin{document}
\title{Schottky Barrier Height Engineering In $\beta$-Ga$_2$O$_3$ Using SiO$_2$ Interlayer Dielectric}
 
\author{Arkka~Bhattacharyya,      
Praneeth Ranga, Muad Saleh, Saurav~Roy, %~\IEEEmembership{Student~Member,~IEEE},
 Michael A. Scarpulla, Kelvin G. Lynn, 
and ~Sriram~Krishnamoorthy\vspace{-0.42cm}
%\thanks{M. Shell was with the Department
%of Electrical and Computer Engineering, Georgia Institute of Technology, Atlanta,
%GA, 30332 USA e-mail: (see http://www.michaelshell.org/contact.html).}% <-this % stops a space
%\thanks{J. Doe and J. Doe are with Anonymous University.}% <-this % stops a space
%\thanks{Manuscript received April 19, 2005; revised August 26, 2015.}
\thanks{Arkka Bhattacharyya, Praneeth Ranga, Saurav Roy, and Sriram Krishnamoorthy are with the Department of Electrical and Computer Engineering, The University of Utah, Salt Lake City, UT, 84112, United States of America (e-mail: a.bhattacharyya@utah.edu, sriram.krishnamoorthy@utah.edu ).}

\thanks{Michael A. Scarpulla is with the Department of Electrical and Computer Engineering and the Department of Materials Science and Engineering, University of Utah, Salt Lake City, Utah 84112, USA.}
\thanks{Muad Saleh is with the Materials Science $\&$ Engineering Program and the Center for Materials Research, Washington State University, Pullman, WA, 99164, United States of America.}
%\thanks{Santosh Swain and Kelvin Lynn is with Center for Materials Research and School of Mechanical and Materials Engineering, Washington State University, Pullman, WA, 99164, United States of America.}
\thanks{Kelvin G. Lynn is with the Materials Science $\&$ Engineering Program, Center for Materials Research, School of Mechanical and Materials Engineering, and the Department of Physics, Washington State University, Pullman, WA, 99164, United States of America.}
}

% The paper headers
\markboth{}%
{Shell \MakeLowercase{\textit{et al.}}: Bare Demo of IEEEtran.cls for IEEE Journals}

% make the title area
\maketitle
\begin{abstract}
This paper reports on the modulation of Schottky barrier heights (SBH) on three different orientations of $\beta$-Ga$_2$O$_3$ by insertion of an ultra-thin SiO$_2$ dielectric interlayer at the metal-semiconductor junction, which can potentially lower the Fermi-level pinning (FLP) effect due to metal-induced gap states (MIGS). Pt and Ni metal-semiconductor (MS) and metal-interlayer-semiconductor (MIS) Schottky barrier diodes were fabricated on bulk n-type doped $\beta$-Ga$_2$O$_3$ single crystal substrates along the (010), (-201) and (100) orientations and were characterized by room temperature current-voltage (I-V) and capacitance-voltage (C-V) measurements. Pt MIS diodes exhibited 0.53 eV and 0.37 eV increment in SBH along the (010) and (-201) orientations respectively as compared to their respective MS counterparts. The highest SBH of 1.81 eV was achieved on the (010)-oriented MIS SBD using Pt metal. The MIS SBDs on (100)-oriented substrates exhibited a dramatic increment ($>$1.5$\times$) in SBH as well as reduction in reverse leakage current. The use of thin dielectric interlayers can be an efficient experimental method to modulate SBH of metal/Ga$_2$O$_3$ junctions.
\end{abstract}
\begin{IEEEkeywords}
gallium oxide, Schottky contact, metal-insulater-semiconductor, Fermi-level pinning, power device
\end{IEEEkeywords}

\IEEEpeerreviewmaketitle
\section{Introduction}
\label{sec1}
\IEEEPARstart{B}{eta} 
-Ga$_2$O$_3$ is a transparent conducting oxide which has emerged as a promising candidate for next generation power electronic devices largely due to its wide band gap (E$_g$ $\sim$ 4.6 - 4.9 eV)\cite{Higashiwaki2012}\cite{He2006}. With a large projected breakdown field of 6-8 MV/cm, the predicted Baliga Figure of Merit (BFOM) is more than three times greater than the conventional wide band gap semiconductors such as SiC and GaN\cite{Higashiwaki2018}. The availability of native single crystal substrates made from cost-effective melt-grown techniques and the ability to grow high quality epitaxial films with controllable doping using advanced epitaxial techniques makes it further attractive for high power vertical devices \cite{Higashiwaki2014, Murakami2015, Sasaki2012, Rafique2016, Shinohara2008, Wagner2014}. However, due to the difficulty with p-type doping and the flat valence band dispersion resulting in very large effective mass for holes, the use of $\beta$-Ga$_2$O$_3$ is currently restricted to unipolar power devices such as metal-semiconductor FETs, MOSFETs and rectifying diodes\cite{He2006, Pearton2018, Stepanov2016}. Schottky contacts with enhanced barrier heights and low reverse leakage currents is crucial for high-power device applications.	Therefore,	the	optimization	of	metal-semiconductor (MS) Schottky contacts (SCs) on $\beta$-Ga$_2$O$_3$ is of key importance for reliable functioning of these unipolar devices. It is of particular interest to investigate whether it is possible to obtain large Schottky barrier heights ($\sim$ E$_g$/2) that can potentially then be used to design Enhancement-mode MESFETs.

\begin{figure}
\centering
  \includegraphics[width=7in,height=7cm,keepaspectratio]{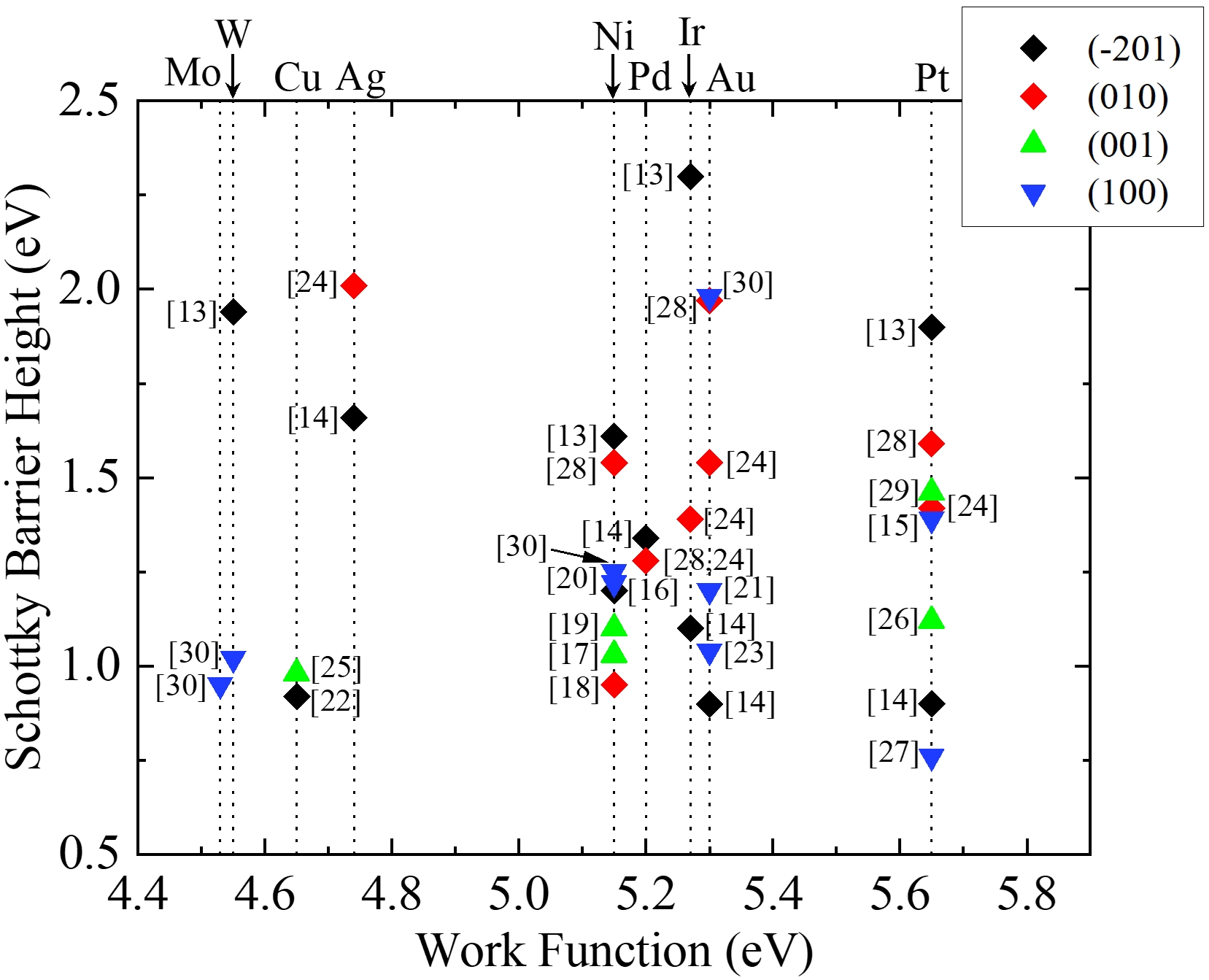}
\label{fig 1}
\caption{Overview of Schottky barrier heights extracted using I-V, C-V and internal photoemission (IPE) measurements on four different orientations of $\beta$-Ga$_2$O$_3$ using different metals as a function of metal workfunction. The metal workfunction values were considered from reference\cite{Michaelson1977}.}
\end{figure}

In the last few years, formation of SCs on $\beta$-Ga$_2$O$_3$ and their electrical properties were studied and investigated, most of which involved SCs with various high workfunction metals, surface treatments and different metal deposition techniques on different orientations of $\beta$-Ga$_2$O$_3$ substrates\cite{Yao2017, Hou2019a, He2017, Armstrong2016, Yang2018, Oh2017, Yang2017, Irmscher2011, Suzuki2009, Splith2014, Mohamed2012, Hou2019, Sasaki2017, Higashiwaki2016, Jian2018, Farzana2017, Konishi2017,Jiang2019}. The anisotropic material properties of $\beta$-Ga$_2$O$_3$ due to its highly asymmetric monoclinic crystal structure has also attracted immense research interest \cite{Pearton2018}, \cite{Stepanov2016}. A brief overview of the measured Schottky barrier heights (SBH) of SCs with high work function metals on various orientations of $\beta$-Ga$_2$O$_3$ is shown in Figure 1. The (010) orientation exhibits lower oxygen-dangling bond density and higher surface band- bending compared to (-201) orientation \cite{Jang2017}, \cite{Fu2018} and is expected to exhibit larger SBH, but Yao et. al.\cite{Yao2017} showed that higher barrier heights can be achieved on (-201) orientation with surface treatments. Farzana et. al.\cite{Farzana2017} reported a range of SBH (1.28-1.97eV) using different metals suggesting that the classical Fermi level pinning effect (FLP) may not be the dominant factor for SC formation on (010) $\beta$-Ga$_2$O$_3$ SBDs, but there are other reports on (010) $\beta$-Ga$_2$O$_3$ with lower reported barrier heights \cite{Oh2017, Hou2019}. Study on (100) and (001) $\beta$-Ga$_2$O$_3$ is rather sparse and till date very low barrier heights have been reported for (100) $\beta$-Ga$_2$O$_3$ \cite{Irmscher2011, Mohamed2012, He2017,Jiang2019,Yang2018,Yang2017,Sasaki2017,Higashiwaki2016}. Furthermore, it is also observed from Figure 1 that the SBH on $\beta$-Ga$_2$O$_3$ does not show an universal trend with the metal workfunction indicating that surface/interface states due to defects and crystal orientation, crystal quality and their passivation with different types of surface treatment or metal deposition techniques can play a very important role in determining the effective SBH.
\begin{figure}
\centering
\includegraphics[width=3.4in,height=4cm,keepaspectratio]{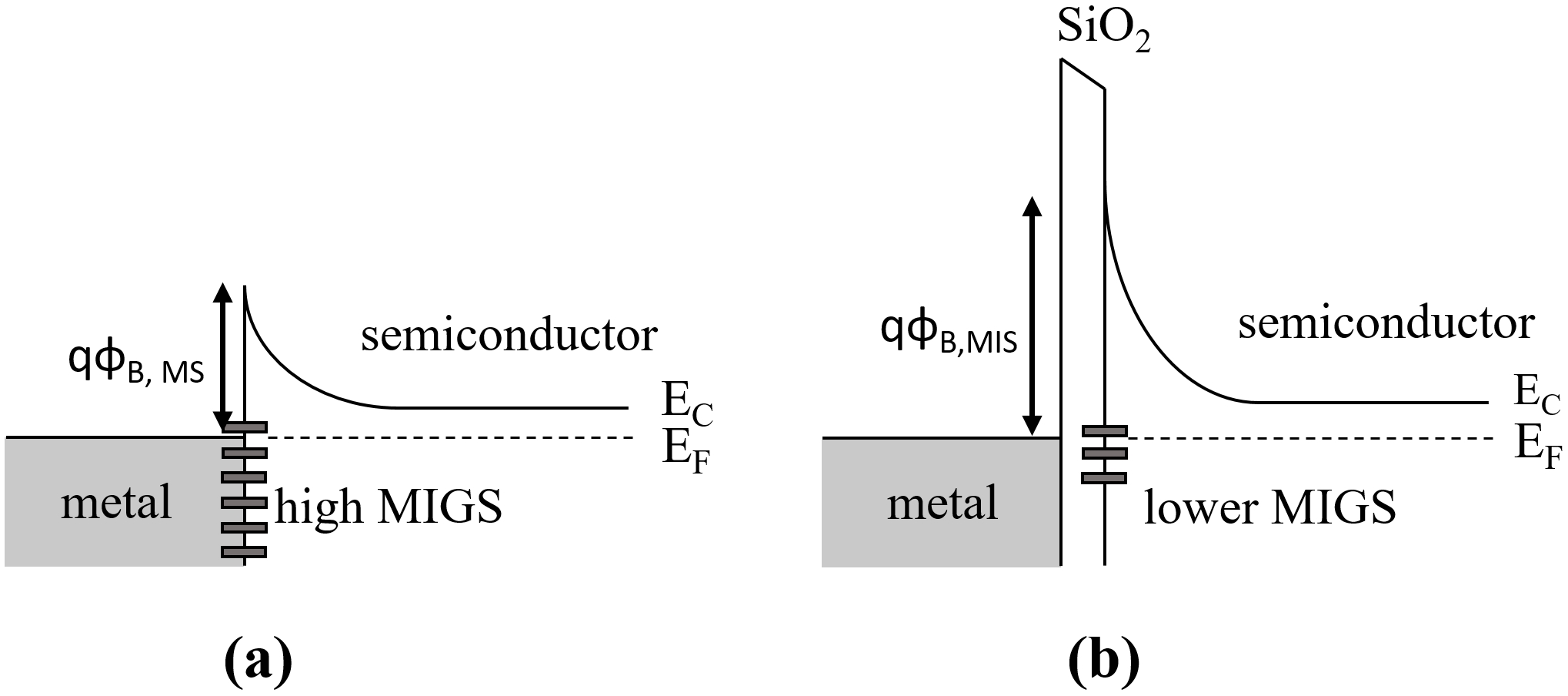}
\caption{Schematic of energy band diagram of (a) MS and (b) MIS Schottky junctions showing the lowering of MIGS with the insertion of SiO$_2$ interfacial layer and a possible enhancement of Schottky barrier height.}
\label{fig2}
\vspace{-0.5cm}
\end{figure}

According to the Schottky-Mott rule, the SBH achieved at a SC is the difference between the metal work function and the semiconductor electron affinity. However, the Schottky-Mott rule is rarely observed. The effective barrier height that is established at a metal-semiconductor interface is actually governed by a combination of various factors such as metal workfunction difference, interface states and the effect of image force lowering \cite{Mohamed2012}. The interface states at a metal-semiconductor junction are mostly mid-gap states that originate from the metal wave functions decaying into the semiconductor band gap and are called metal-induced gap states (MIGS) \cite{Robertson2006}. The other contribution to the interface states come from the reconstruction of the dangling bonds, defects and localized impurities at the metal-semiconductor interface \cite{Monch1999}. Depending on the density of these interface states, the Fermi level gets pinned near one of the band edges and thus play a very important role in determining the effective barrier height that can be measured. The weak dependence of SBH on the metal workfunction has also been observed and studied in other semiconductor materials like Ge, Si, and InGaAs and is attributed to FLP caused by metal-induced gap states or defects at the metal-semiconductor interface \cite{Roy2010, Lin2011, Agrawal2012, Agrawal2014, Monch1999}. Many groups in the past have reported that the introduction of a thin interfacial dielectric layer, both in-situ and ex-situ, can act as a blocking layer to prevent the spilling of metal electron waves and thus can potentially lower the FLP effect due to MIGS \cite{Roy2010, Lin2011, Agrawal2012, Agrawal2014} (Fig 2(b)). This provides a simple and elegant solution to engineer the effective barrier height by reducing the contribution from MIGS. In this work, we investigate the modulation of Schottky barrier height on different orientations of $\beta$-Ga$_2$O$_3$ single crystal substrates with the insertion of ultra-thin SiO$_2$ dielectric layer at the metal-semiconductor interface.

\section{Device Fabrication and Characterization}
\label{sec1a}
The  5 mm  $\times$  5 mm  $\times$  0.6 mm  edge-defined film-fed grown (EFG) Sn-doped
(010) and (-201) $\beta$-Ga$_2$O$_3$ substrates were acquired from Novel Crystal Tech (Japan). The Zr-doped (100) $\beta$-Ga$_2$O$_3$  single crystal bulk substrates were grown by vertical gradient freeze (VGF) method and the details are available in reference \cite{Saleh2019}. The (100)-oriented samples were prepared by sawing first and then cleaving along the cleavage plane (100) into samples of  3.5 $\times$ 4.5 $\times$ 0.6 mm$^3$ dimensions and the substrate orientation was confirmed by XRD measurements and reported elsewhere \cite{Saleh2019}. On (010) oriented substrates, the electron concentration and mobility from Hall measurements were measured to be 1.1$\times$10$^{18}$ cm$^{-3}$ and 89 cm$^{2}$/Vs, respectively. For the (-201) oriented substrates, the electron concentration and mobility values measured were 1.7$\times$10$^{18}$ cm$^{-3}$ and 32 cm$^{2}$/Vs, respectively. From Hall effect measurements, the room temperature net electron concentration and mobility were measured to be 1.2$\times$10$^{18}$ cm$^{-3}$ and 78 cm$^{2}$/Vs, respectively in the (100)-oriented samples. It should be noted that the electron concentration is similar for the samples along all the three orientations considered here for this study. The electron concentrations and doping profile were also further confirmed using capacitance-voltage measurements as discussed later in the paper. 

Six substrates, two of each orientation, were first cleaned using conventional solvents (acetone, IPA and DI water) followed by dip in Piranha solution (98$\%$ H$_2$SO$_4$: 32$\%$ H$_2$O$_2$ 4:1) for 5 mins. Three substrates, one of each orientation, were processed as
\begin{figure}
\centering
\includegraphics[width=7.5in,height=7.5cm,keepaspectratio]{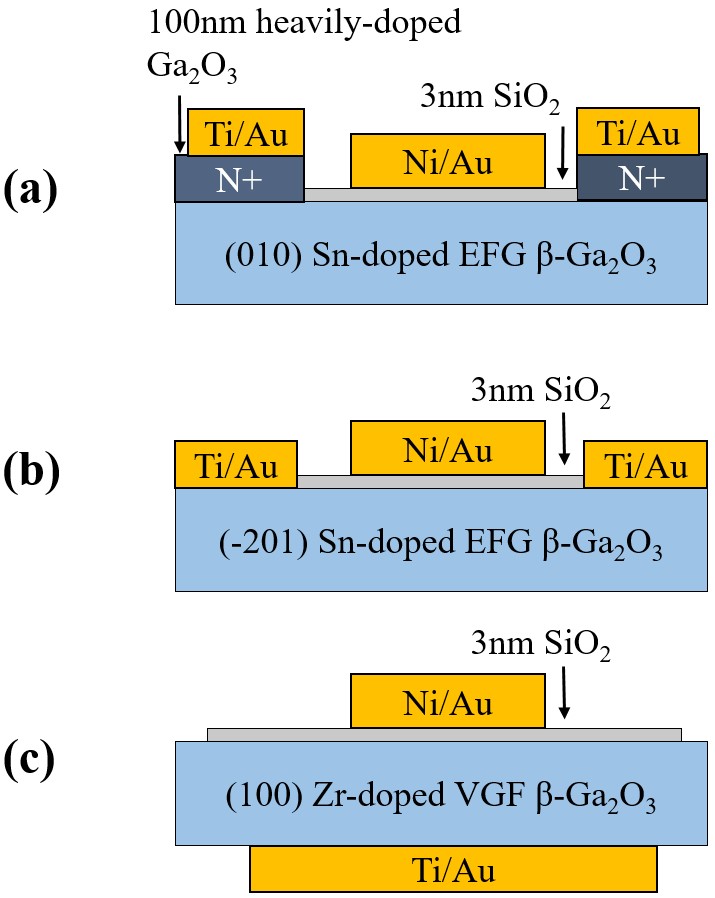}
\caption{Schematic of MIS diode structures on (a) (010), (b) (-201), and (100) oriented $\beta$-Ga$_2$O$_3$ substrates with 3 nm SiO$_2$ interlayer using Ni Schottky metal (Pt MIS diodes had identical structures with Pt as anode instead of Ni).  The respective MS diodes are similar in structure just without the SiO$_2$ dielectric interlayer.}
\label{fig3}
\vspace{-0.1cm}
\end{figure}
MS diodes and the rest three substrates, one of each orientation, were processed as metal-interlayer (SiO$_2$)-semiconductor (MIS) diodes. Series resistance effect was dominant in the capacitance voltage measurements on the (010) and (-201) SBDs necessitating formation of quasi-lateral diodes with concentric Ohmic-Schottky design with 5-30 $\mu$m spacing between the Ohmic and Schottky pads. For the (010) oriented substrate, first an extra step of heavily-doped Ga$_2$O$_3$ (100 nm thick, $N_D$(Si) $\sim$ 1$\times$10$^{20}$ cm$^{-3}$) was selectively grown in the ohmic contact regions by Agnitron Agilis MOCVD system using 500 nm thick SiO$_2$ (PECVD) masks to realize good ohmic contacts. Then Ti/Au (50 nm/50 nm) was sputtered in the Ohmic contact regions defined by photolithography and lift-off process followed by rapid thermal annealing at 450$^o$C in nitrogen for 1.5 minutes. On the (-201) oriented substrates, first Ti/Au (50 nm/50 nm) Ohmic contacts were sputter deposited and patterned using photolithography and lift-off process and no further processing was needed to realize good ohmic contacts. Following this, SiO$_2$ dielectric was deposited by ALD (discussed in the next paragraph) on the (010) and (-201) MIS samples and the oxide in the contact region was etched using a quick dip (10 seconds) in diluted HF solution after patterning by standard optical lithography. Next, 150 $\mu$m and 200 $\mu$m diameter circular Pt/Au (50 nm/50 nm) and Ni/Au (50 nm/50 nm) Schottky contacts were sputtered and e-beam evaporated respectively on the MS and MIS samples (both (010) and (-201)) after re-aligning to the ohmic contacts using standard photolithography. For the (100) oriented samples, SiO$_2$ dielectric was first deposited by ALD on the front side of MIS sample and then Ti/Au (50nm/50nm) ohmic contacts were sputtered on the backside of the sample. Then 150 $\mu$m and 200 $\mu$m diameter Pt/Au (50nm/50nm) and Ni/Au (50nm/50nm) Schottky contacts were sputter deposited and e-beam evaporated respectively on both MS and MIS samples. The MIS diodes on all three substrates were not subjected to any high temperature process after the ALD dielectric deposition. The processed MIS diode schematics are shown in Figure 3. The current-voltage (I-V) characteristics and capacitance-voltage (C-V) measurements (1 MHz) were performed in air at room temperature ($\sim$298K) using a Keithley 4200A-SCS parameter analyzer.

Before loading the MIS samples into the ALD chamber, they were first solvent cleaned (acetone, IPA and DI water) followed by dip in Piranha solution (98$\%$ H$_2$SO$_4$: 32$\%$ H$_2$O$_2$ 4:1) for 5 mins. Before the start of the ALD deposition cycle, the substrates were treated with remote oxygen plasma (300W and 20 sccm O$_2$ flow) for 5 minutes. A 3nm thin SiO$_2$ layer was deposited on the three substrates for MIS processing at 200$^o$C using a Cambridge Fiji F200 ALD tool using tris(dimethylamino)silane (3DMAS) precursor and O$_2$ plasma source. The oxide thickness was confirmed by performing optical ellipsometry on a monitor Si wafer using a Woollam V-VASE spectroscopic ellipsometer tool. The measured thickness of SiO$_2$ layer was ~3.5 nm on the Si wafer and the SiO$_2$ formed on the Si wafer due to the remote plasma treatment was measured to be 4-5 \AA. SiO$_2$ thickness on  Ga$_2$O$_3$ is hence estimated to be 3 nm, and this is used as the interlayer thickness for further analysis. 

\begin{figure}
\centering
\begin{subfigure}
  \centering
  \includegraphics[width=6in,height=6cm,keepaspectratio]{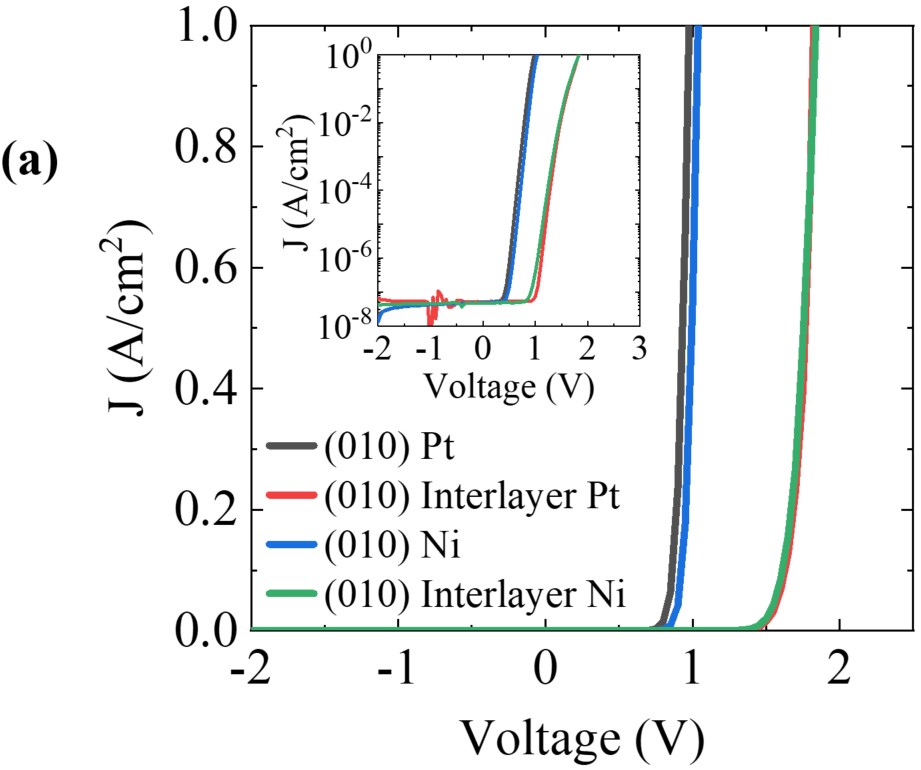}
  \label{fig1:sub1}
\end{subfigure}
\begin{subfigure}
  \centering
  \includegraphics[width=6in,height=6cm,keepaspectratio]{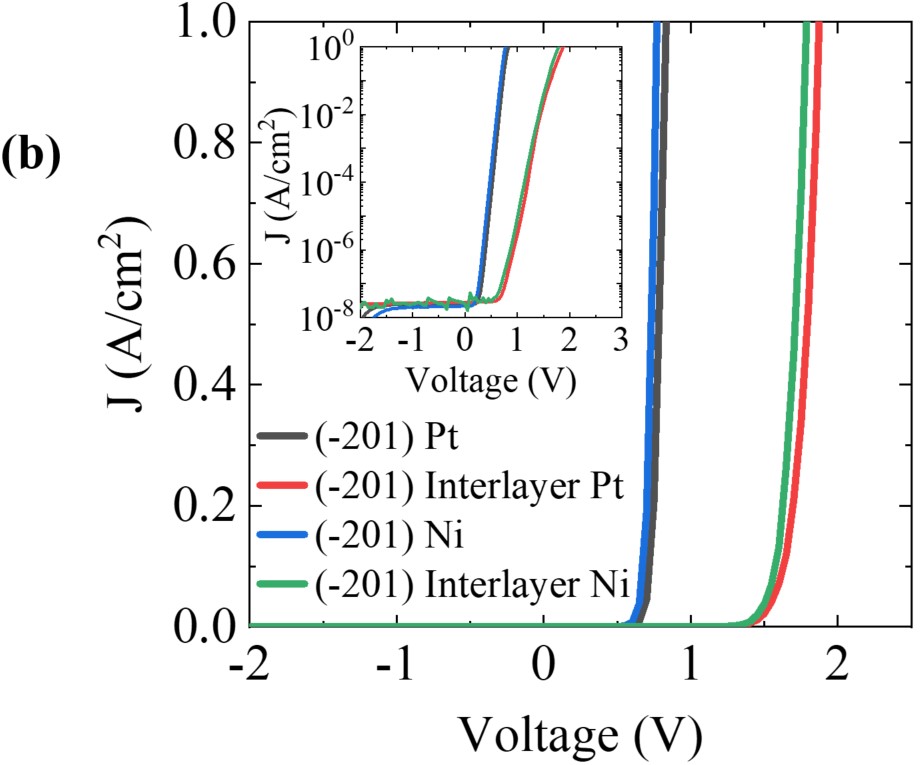}
  \label{fig1:sub2}
\end{subfigure}
\begin{subfigure}
  \centering
  \includegraphics[width=6in,height=6cm,keepaspectratio]{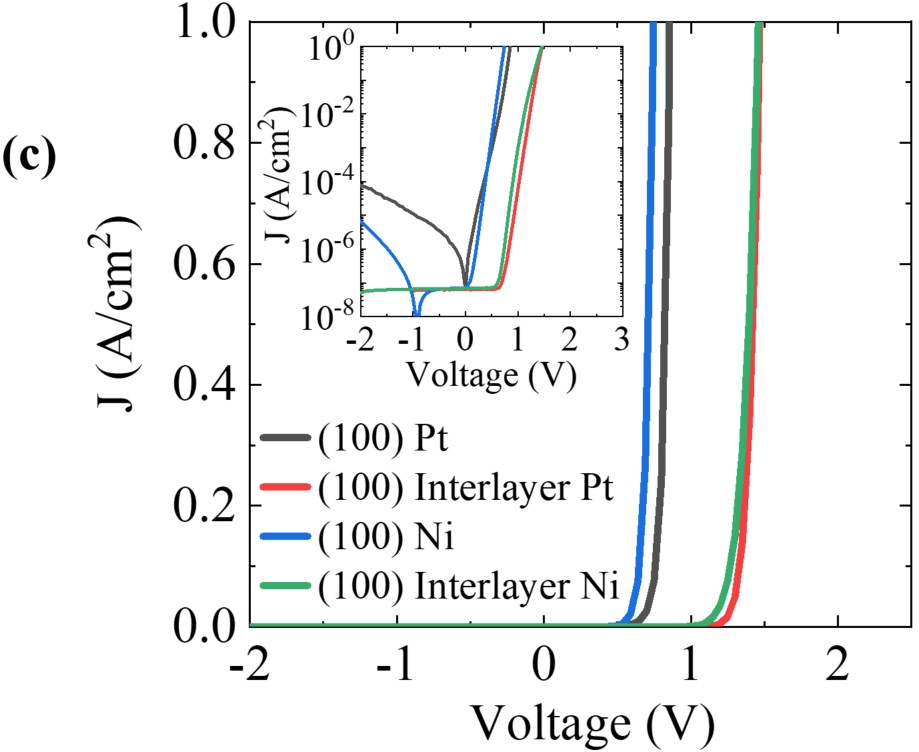}
  \label{fig1:sub3}
\end{subfigure}
\label{fig 4}
\caption{Linear J-V characteristics (200 $\mu$m diameter pad size) of the Pt and Ni MS and MIS SBDs on (a) (010), (b) (-201) and (c) (100) - oriented $\beta$-Ga$_2$O$_3$ showing the increase in forward voltages with the insertion of ultra-thin SiO$_2$ and the insets showing their corresponding log J-V characteristics.}
\vspace{-0.5cm}
\end{figure} 

\section{RESULTS AND DISCUSSIONS}
\label{sec2}

\begin{table*}[!htb]
\captionsetup{size=footnotesize}
\caption{SUMMARY OF EXTRACTED SBH FROM J-V CHARACTERISTICS FOR ALL MS AND MIS SBDs USING TE MODEL} \label{tab1}
\setlength\tabcolsep{0pt} % let LaTeX compute intercolumn whitespace
\smallskip 
\begin{tabular*}{\textwidth}{@{\extracolsep{\fill}}rcccccr}
\hline
\vspace{0.03cm} &\vspace{0.03cm}&\vspace{0.03cm}&\vspace{0.03cm}&\vspace{0.03cm}&\vspace{0.03cm}&\vspace{0.03cm}\\
%\toprule
  Substrate & Metal & $q\Phi^{IV}_{B, MS}$ (eV) & $n_{MS}$  & $q\Phi^{IV}_{B, MIS}$ (eV) & $n_{MIS}$ & $\Delta q\Phi^{IV}_B$ (eV)\\
  \vspace{0.03cm} &\vspace{0.03cm}&\vspace{0.03cm}&\vspace{0.03cm}&\vspace{0.03cm}&\vspace{0.03cm}&\vspace{0.03cm}\\
   \hline
%\midrule
 \vspace{0.03cm} &\vspace{0.03cm}&\vspace{0.03cm}&\vspace{0.03cm}&\vspace{0.03cm}&\vspace{0.03cm}&\vspace{0.03cm}\\
 \multirow{2}{*}{(010)}  & Pt & 1.18 $\pm$ 0.06 & 1.15 $\pm$ 0.08 & 1.56 $\pm$ 0.08 & 1.36 $\pm$ 0.1 & 0.38 $\pm$ 0.14\\
                         & Ni & 1.27 $\pm$ 0.04 & 1.16 $\pm$ 0.05 & 1.38 $\pm$ 0.06 & 1.61 $\pm$ 0.1 & 0.11 $\pm$ 0.1\\
 \vspace{0.01cm} &\vspace{0.01cm}&\vspace{0.01cm}&\vspace{0.01cm}&\vspace{0.01cm}&\vspace{0.01cm}&\vspace{0.01cm}\\
 \hline
  \vspace{0.01cm} &\vspace{0.01cm}&\vspace{0.01cm}&\vspace{0.01cm}&\vspace{0.01cm}&\vspace{0.01cm}&\vspace{0.01cm}\\
 \multirow{2}{*}{(-201)} & Pt & 1.08 $\pm$ 0.08 & 1.13 $\pm$ 0.2 & 1.30 $\pm$ 0.05 & 1.37 $\pm$ 0.1 & 0.22 $\pm$ 0.13\\
                        & Ni & 1.04 $\pm$ 0.04 & 1.15 $\pm$ 0.08 & 1.21 $\pm$ 0.07 & 1.91 $\pm$ 0.2 & 0.17 $\pm$ 0.1\\
  \vspace{0.01cm} &\vspace{0.01cm}&\vspace{0.01cm}&\vspace{0.01cm}&\vspace{0.01cm}&\vspace{0.01cm}&\vspace{0.01cm}\\
 \hline
  \vspace{0.01cm} &\vspace{0.01cm}&\vspace{0.01cm}&\vspace{0.01cm}&\vspace{0.01cm}&\vspace{0.01cm}&\vspace{0.01cm}\\
  \multirow{2}{*}{(100)} & Pt & 0.84 $\pm$ 0.03 & 1.56 $\pm$ 0.07& 1.22 $\pm$ 0.04 & 1.66 $\pm$ 0.08 & 0.38 $\pm$ 0.07\\
                          & Ni & 0.72 $\pm$ 0.02 & 1.51 $\pm$ 0.04 & 1.24 $\pm$ 0.03 & 1.41 $\pm$ 0.05 & 0.52 $\pm$ 0.05\\
  \vspace{0.01cm} &\vspace{0.01cm}&\vspace{0.01cm}&\vspace{0.01cm}&\vspace{0.01cm}&\vspace{0.01cm}&\vspace{0.01cm}\\
 \hline
 \vspace{0.01cm} &\vspace{0.01cm}&\vspace{0.01cm}&\vspace{0.01cm}&\vspace{0.01cm}&\vspace{0.01cm}&\vspace{0.01cm}\\
 % \midrule
%\bottomrule
\end{tabular*}
\footnotesize
$q\Phi^{IV}_{B, MS}$, $q\Phi^{IV}_{B, MIS}$ = Schottky barrier heights (eV) and $n_{MS}$, $n_{MIS}$ = ideality factors of MS and MIS diodes respectively from J-V characteristics. $\Delta q\Phi^{IV}_B$ = $q\Phi^{IV}_{B, MIS}$ - $q\Phi^{IV}_{B, MS}$ (eV).
\end{table*}
The current density-voltage (J-V) characteristics of all the representative Schottky diodes at RT are shown in Figure 4. Both the metal-semiconductor (MS) and metal-interlayer-semiconductor (MIS) Schottky diodes exhibited highly rectifying behavior with $>$ 8 orders of magnitude of rectification at $\pm$2V along the (010) and (-201) orientations (Fig. 4 (a), (b)). The MS diodes on (100) substrates (Fig. 4(c)) were found to be less rectifying. The MIS SBDs showed an increased forward voltage compared to the MS diodes, along all the orientations as expected, indicating that the SBH of MIS diodes might be higher than their respective bare metal MS counterparts, in addition to the blocking of current due to the band offset at the SiO$_2$/Ga$_2$O$_3$ interface with the insertion of an insulator\cite{Jia2015}.

For moderately-doped semiconductors, generally, thermionic emission (TE) is the dominant transport mechanism in ideal MS diodes \cite{Sze2006}. The J-V characteristics of the MS SBDs and MIS SBDs were analyzed using the TE model which can be expressed as

\begin{equation}
J = A^{**}T^2e^{-\frac{q\phi_B^{eff}}{kT}}\left(e^{\frac{qV}{nkT}}-1\right)
\label{eq1}
\end{equation}
\vspace{-0.3cm} 
where,
\begin{equation}
J_o = A^{**}T^2e^{-\frac{q\phi_B^{eff}}{kT}}
\label{eq2}
\end{equation}
where $A^{**}$ is the effective Richardson constant, with  a calculated theoretical value of 41.1 A cm$^{-2}$K$^{-2}$ (for electron effective   mass  of $m_e^* = 0.34m_o$  )\cite{Pearton2018}, $q$ is the elementary charge, $k$ is the Boltzmann constant, $V$ is applied bias voltage, $n$ is the ideality factor, $\Phi^{eff}_B$ is the effective barrier height, $J_o$ is the reverse saturation current density, and $T$ is the absolute temperature. The effective barrier height is then calculated as,
\begin{equation}
q\Phi_B^{eff} = kTln\left(\frac{A^{**}T^2}{J_0}\right)
\label{eq3}
\end{equation}
\vspace{-0.1cm} 
and the ideality factor, $n$, is defined as,
\begin{equation}
n = \frac{q}{2.3kT\left(\frac{dlog(J)}{dV}\right)}
\label{eq4}
\end{equation}
\vspace{-0.3cm} 

The barrier heights and ideality factors extracted from the J-V characteristics are summarized in Table I. The barrier heights for the MS SBDs were in the range 0.72 eV to 1.27 eV with lowest value for Ni on (100) substrate and the highest for Pt on (010) substrate. The extracted SBH values are comparable to most reports in the literature (Figure 1). The MS Pt and Ni diodes on the (100) oriented substrate exhibited lower barrier heights with higher values of $n$ than the other two orientations which indicates higher degree of contribution from non-thermionic transport mechanisms. This effect has been observed in other reports on floating zone (FZ), Czochralski (CZ) and EFG grown (100) $\beta$-Ga$_2$O$_3$ bulk crystals \cite{Suzuki2009, Mohamed2012, He2018}. For the MIS SBDs, the extracted SBH were in a range of 1.21 eV to 1.56 eV with Ni on (-201) being the lowest and Pt on (010) being the highest. Although, this may indicate an improvement in barrier heights with the insertion of an SiO$_2$ interlayer, but still these values are an underestimation as we will see in subsequent discussions. TE model can underestimate the barrier heights for non-ideal diodes (n$>$1) due to barrier height inhomogeneities at the MS junction \cite{Sze2006, Werner1991}.

The  MIS  SBDs  on  all  the  orientations  exhibited comparatively higher  $n$ values which is expected and also  has been observed in previously published reports in other semiconductor   systems   \cite{Card1971, Cowley1966}. The   presence   of   an intentional or unintentional interfacial layer could result in tunneling of electrons through the insulator and enhanced surface band bending at the dielectric-semiconductor interface. Solving the metal-oxide semiconductor electrostatics taking into account the voltage drop across the thin oxide and also the interface trap charge, the ideality factor for non-ideal MIS Schottky diodes on n-type semiconductor can be modeled as a function of interface density of trap states, D$_{it}$ and also the interfacial layer thickness as done by Card and Rhoderick \cite{Card1971},

\begin{equation}
n = 1 + \frac{\delta}{\epsilon_{ox}}\left(\frac{\epsilon_s}{W}+qD_{it}\right)
\label{eq5}
\end{equation}

where, $n$ is the ideality factor extracted from the TE model, $\delta$ is the interlayer oxide thickness, $\epsilon_{ox}$ is the permittivity of the oxide, $\epsilon_{s}$ is the semiconductor permittivity, $W$ is the depletion depth inside the semiconductor, $q$ is the elementary charge and $D_{it}$ is the interface state density. This model, although, not very accurate when the interface state densities are very high, it can be very effective for estimation of mean D$_{it}$ value, especially for ultra-thin oxides when conventional C-V measurement techniques, such as high-low method, quasi-static measurements become unviable because of very high dissipation losses even at very low forward bias while the device is moved from depletion to accumulation. The dual sweep I-V characteristics (-3V to 3V to -3V) of the MIS SBDs show very low hysteresis for the (100), (010) and (-201)-oriented substrates indicating minimal charge trapping at the semiconductor-dielectric interface. However, the (-201) MIS SBDs exhibited comparably a little higher hysteresis than other two orientations which can be attributed to the presence of higher $D_{it}$ as previously reported \cite{Zeng2017}. Nevertheless, all the MIS diodes exhibited low hysteresis ($\Delta$V $<$ 0.15V) indicating good quality interfaces for the MIS SBDs. Hence, we use the measured value of $n$ to estimate D$_{it}$ in the MIS diodes. 

SBH  values  extracted  from J-V  characteristics  in general,
can underestimate the barrier height because of the barrier height inhomogeneity and current conduction through localized low SBH regions. We performed C-V measurements on both the MS and  MIS diodes along the (010), (100) and (-201) orientations. First, we assume that the voltage drop across thin dielectric SiO$_2$ interfacial layer to be negligible. For a Schottky-diode under bias, the C-V relationship can be expressed as \cite{Sze2006},

%\begin{equation}
%W = \sqrt{\frac{2\epsilon_s}{qN_D}\left(V_{bi} - V - \frac{kT}{q}\right)}
%\label{eq1}
%\end{equation}
%\vspace{-0.3cm} 

% From this, the semiconductor capacitance can be expressed as,

\begin{equation}
C = \frac{A\epsilon_s}{W} = A\sqrt{\frac{q\epsilon_sN_D}{2\left(V_{bi} - V - \frac{kT}{q}\right)}}
\label{eq6}
\end{equation}
\vspace{-0.3cm} 

and
\vspace{-0.1cm}
\begin{equation}
\frac{A^2}{C^2} = \frac{2\left(V_{bi} - V - \frac{kT}{q}\right)}{q\epsilon_sN_D}
\label{eq7}
\end{equation}

where, $\epsilon_s$  is the semiconductor permittivity (for $\beta$-Ga$_2$O$_3$, $\epsilon_s$ = 10$\epsilon_o$ \cite{Pearton2018}, where $\epsilon_o$ = permittivity of free space), $V_{bi}$  is the built-in potential, $N_D$ is the doping concentration in the semiconductor, $A$ is the area of the anode and $W$ is the semiconductor depletion width. $V_{bi}$ and $N_D$ can be extracted from the V-axis intercept and the slope of (A/C)$^{2}$-V plots respectively. Figure 5 shows the room temperature C-V (inset) and (A/C)$^{2}$-V plots of all the SBDs measured at 1MHz. Any variations in the (A/C)$^{2}$-V slopes can be attributed to slight fluctuation in the doping for various Ga$_2$O$_3$ substrates used in this work. However, the doping profiles were flat (Figure 5(d)) for all the three orientations and the net electron concentrations were similar ($\sim1\times$10$^{18}$ cm$^{-3}$) and matched with the Hall measurements. The barrier height is then extracted using the expression,
\begin{figure}
\centering
\begin{subfigure}
  \centering
  \includegraphics[width=5.2in,height=5.2cm,keepaspectratio]{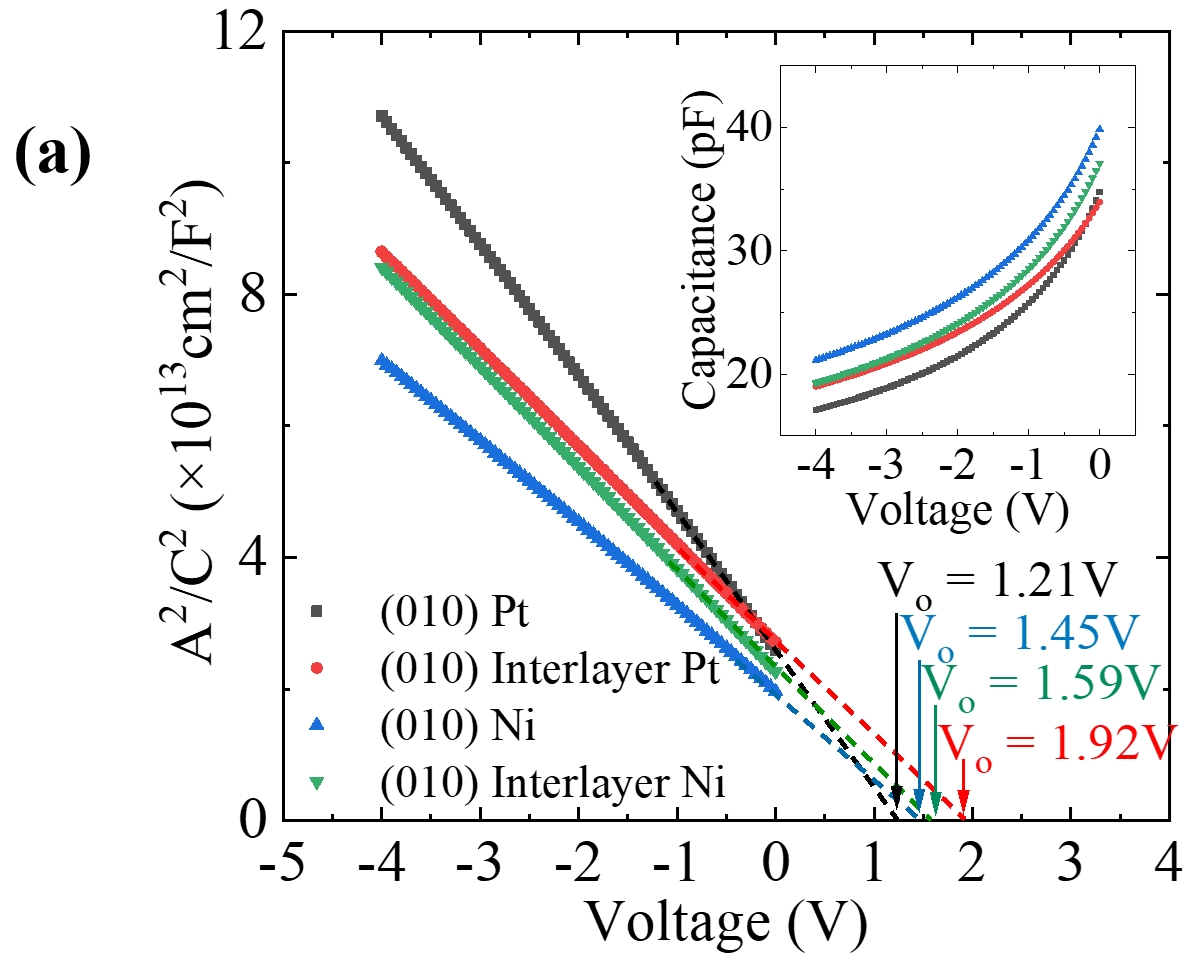}
  \label{fig5:sub1}
\end{subfigure}
\begin{subfigure}
  \centering
  \includegraphics[width=5.2in,height=5.2cm,keepaspectratio]{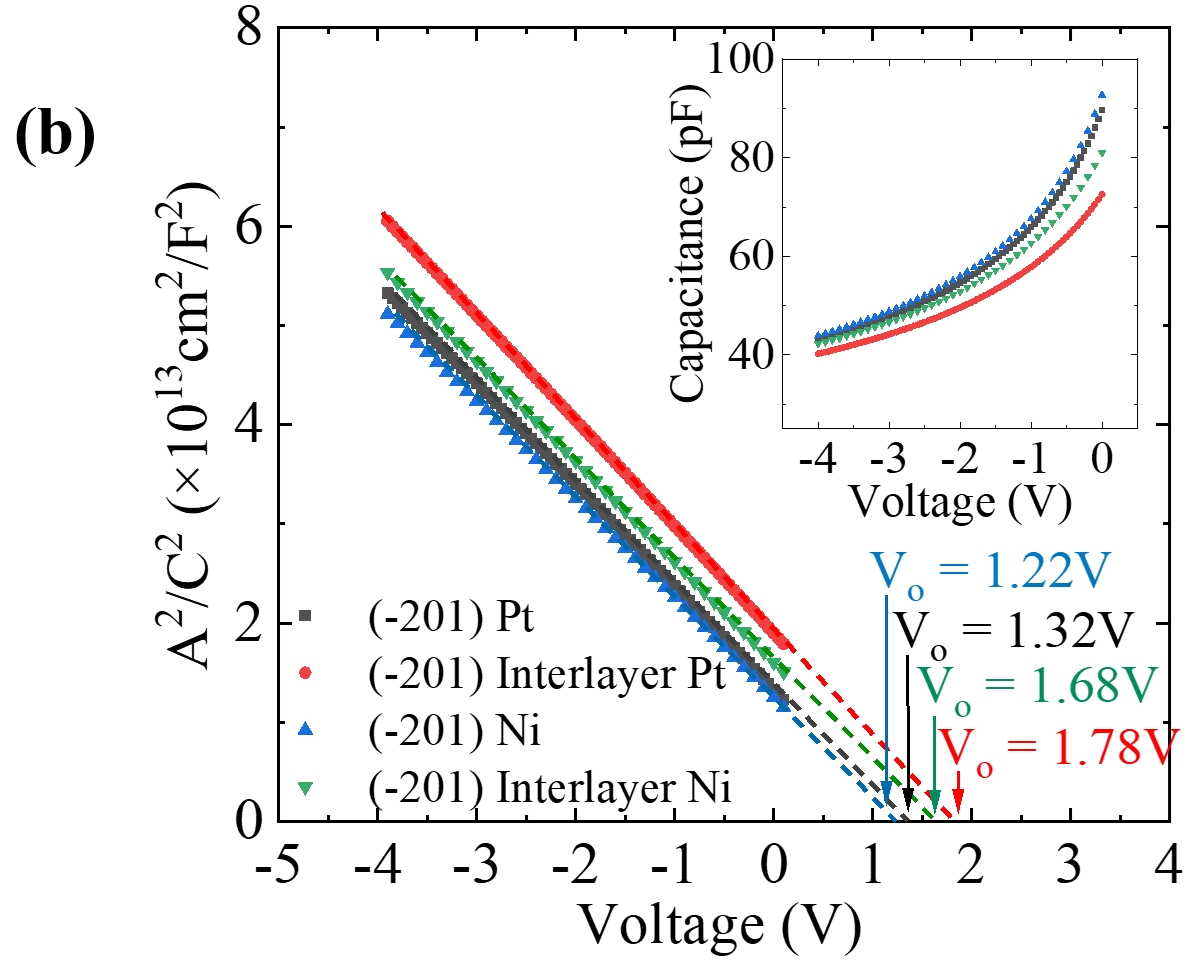}
  \label{fig5:sub2}
\end{subfigure}
\begin{subfigure}
  \centering
  \includegraphics[width=5.2in,height=5.2cm,keepaspectratio]{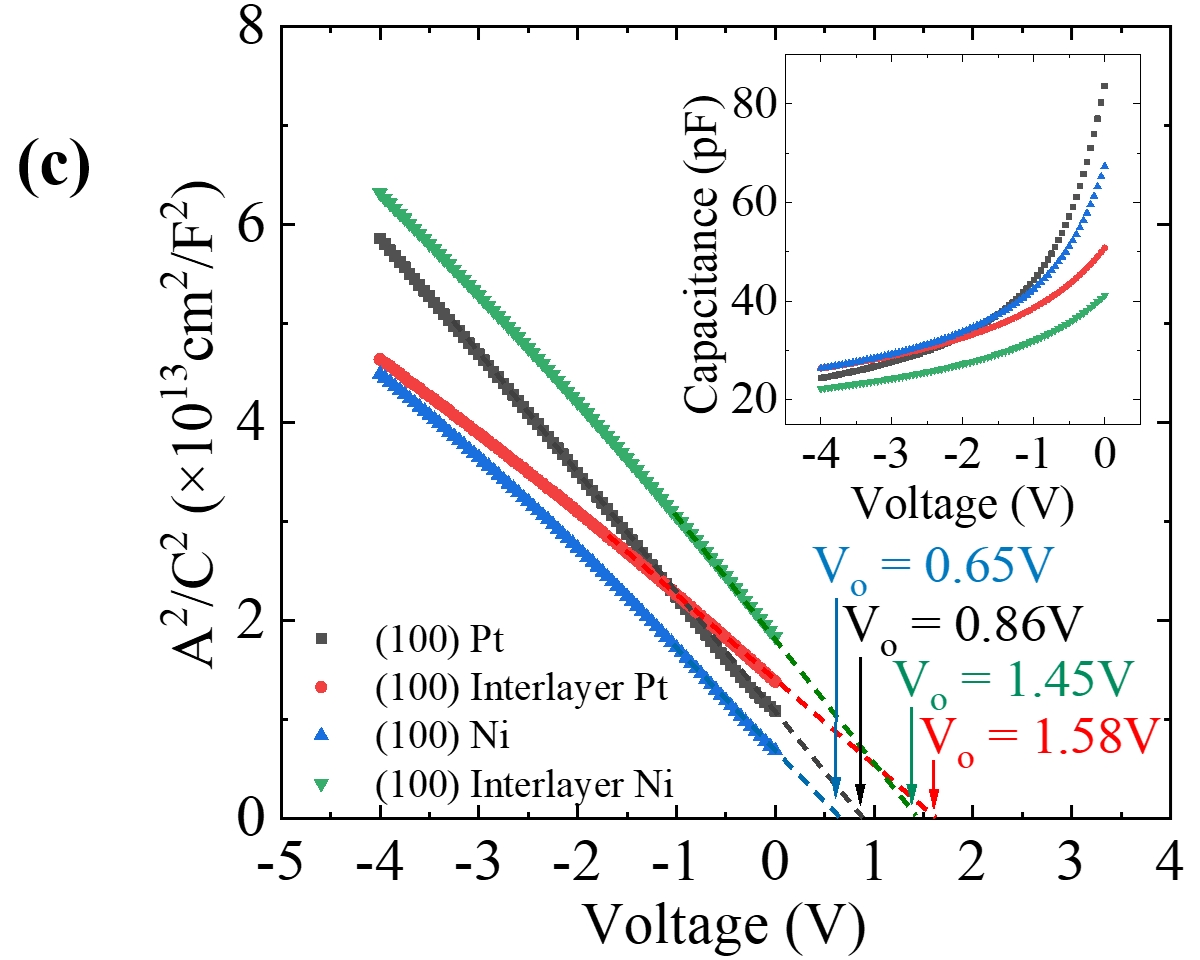}
  \label{fig5:sub3}
\end{subfigure}
\begin{subfigure}
  \centering
  \includegraphics[width=5.2in,height=5.2cm,keepaspectratio]{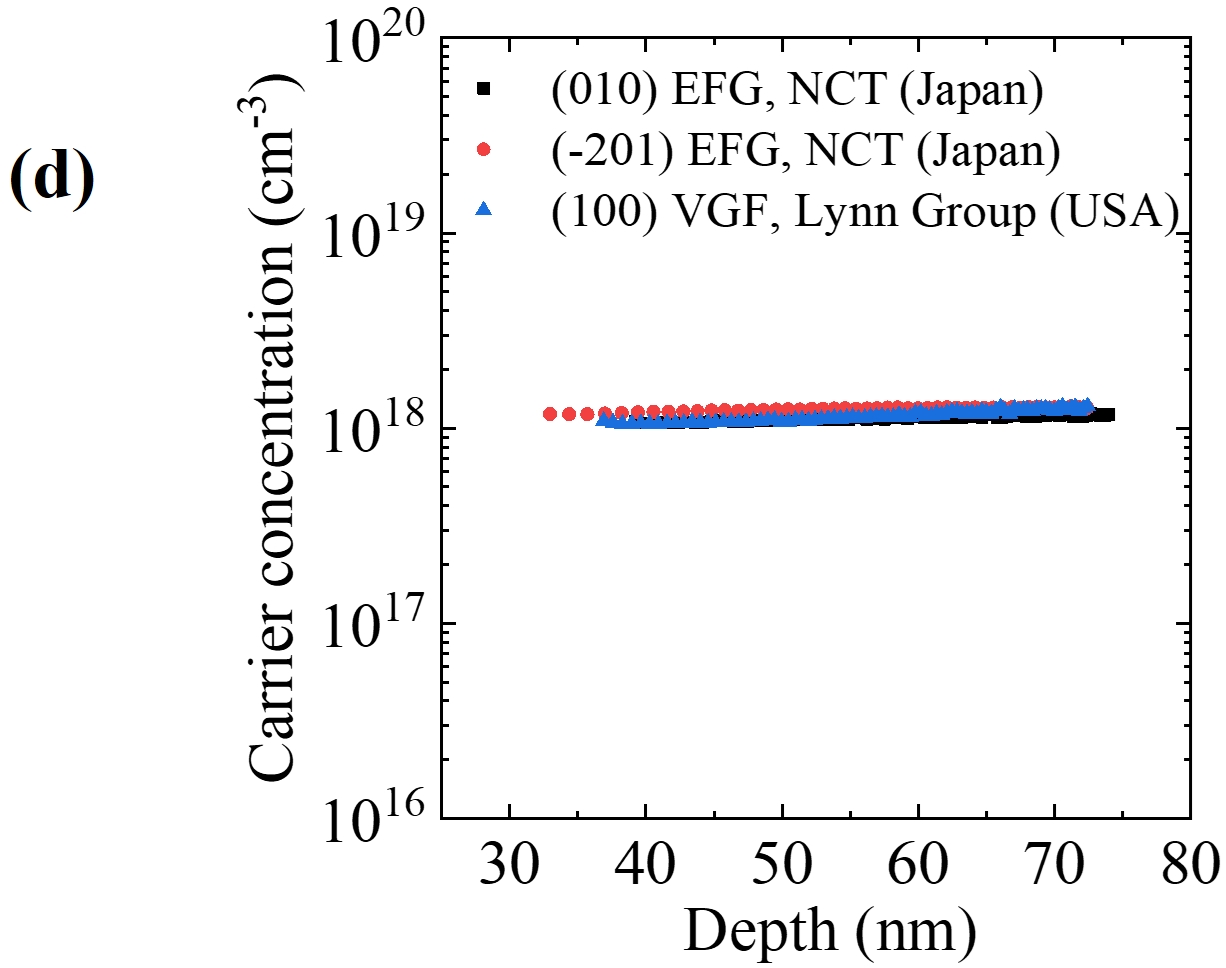}
  \label{fig5:sub4}
\end{subfigure}
\label{fig 5}
\caption{(A/C)$^{2}$-V characteristics of the Pt and Ni MS and MIS SBDs on (a) (010), (b) (-201) and (c) (100) - oriented $\beta$-Ga$_2$O$_3$ showing the increment in intercept voltages with the insertion of ultra-thin SiO$_2$ and the insets showing their corresponding C-V characteristics. (d) Net carrier concentration vs depth profile obtained from C-V measurements of representative devices along three orientations of $\beta$-Ga$_2$O$_3$ substrates used in this work.((010) EFG : 1.02 $\pm$ 0.3$\times$10$^{18}$ cm$^{-3}$, (-201) EFG : 1.6 $\pm$ 0.1$\times$10$^{18}$ cm$^{-3}$, (100) VGF : 1.4 $\pm$ 0.2$\times$10$^{18}$ cm$^{-3}$)}
\vspace{-0.5cm}
\end{figure}

\begin{table*}[!htb]
\captionsetup{size=footnotesize}
\caption{SUMMARY OF EXTRACTED SBH FROM C-V CHARACTERISTICS FOR ALL MS AND MIS SBDs} \label{tab2}
\setlength\tabcolsep{0pt} % let LaTeX compute intercolumn whitespace

\smallskip 
\begin{tabular*}{\textwidth}{@{\extracolsep{\fill}}rcccccr}
\hline
\vspace{0.01cm} &\vspace{0.01cm}&\vspace{0.01cm}&\vspace{0.01cm}&\vspace{0.01cm}&\vspace{0.01cm}&\vspace{0.01cm}\\
%\toprule
  Substrate & Metal & $q\Phi^{CV}_{B, MS}$ (eV) & $q\Phi^{CV}_{B, MIS}$ (eV) & $q\Phi^{CV,\delta}_{B, MIS}$ (eV) & $\Delta q\Phi^{CV}_B$ (eV)\\
  \vspace{0.01cm} &\vspace{0.01cm}&\vspace{0.01cm}&\vspace{0.01cm}&\vspace{0.01cm}&\vspace{0.01cm}&\vspace{0.01cm}\\
%\midrule
\hline
 \vspace{0.03cm} &\vspace{0.03cm}&\vspace{0.03cm}&\vspace{0.03cm}&\vspace{0.03cm}&\vspace{0.03cm}\\
 \multirow{2}{*}{(010)}  & Pt & 1.28 $\pm$ 0.04 & 1.97 $\pm$ 0.05 & 1.81 $\pm$ 0.05& 0.53 $\pm$ 0.09\\
                         & Ni & 1.50 $\pm$ 0.03 & 1.64 $\pm$ 0.02 & 1.54 $\pm$ 0.02 & 0.04 $\pm$ 0.05\\
 \vspace{0.01cm} &\vspace{0.01cm}&\vspace{0.01cm}&\vspace{0.01cm}&\vspace{0.01cm}&\vspace{0.01cm}\\
 \hline
  \vspace{0.01cm} &\vspace{0.01cm}&\vspace{0.01cm}&\vspace{0.01cm}&\vspace{0.01cm}&\vspace{0.01cm}&\\
  \multirow{2}{*}{(-201)} & Pt & 1.38 $\pm$ 0.08 & 1.84 $\pm$ 0.03 & 1.75 $\pm$ 0.03 & 0.37 $\pm$ 0.11\\
                         & Ni & 1.26 $\pm$ 0.05 & 1.74 $\pm$ 0.06 & 1.28 $\pm$ 0.06 & 0.02 $\pm$ 0.11\\
  \vspace{0.01cm} &\vspace{0.01cm}&\vspace{0.01cm}&\vspace{0.01cm}&\vspace{0.01cm}&\vspace{0.01cm}\\
 \hline
  \vspace{0.01cm} &\vspace{0.01cm}&\vspace{0.01cm}&\vspace{0.01cm}&\vspace{0.01cm}&\vspace{0.01cm}\\
  \multirow{2}{*}{(100)} & Pt & 0.92 $\pm$ 0.02& 1.63 $\pm$ 0.03 & 1.44 $\pm$ 0.03 & 0.52 $\pm$ 0.04\\
                          & Ni & 0.71 $\pm$ 0.02 & 1.51 $\pm$ 0.02 & 1.32 $\pm$ 0.02& 0.61 $\pm$ 0.05 \\
  \vspace{0.01cm} &\vspace{0.01cm}&\vspace{0.01cm}&\vspace{0.01cm}&\vspace{0.01cm}&\vspace{0.01cm}\\
 \hline
 \vspace{0.01cm} &\vspace{0.01cm}&\vspace{0.01cm}&\vspace{0.01cm}&\vspace{0.01cm}&\vspace{0.01cm}\\
 % \midrule
%\bottomrule
\end{tabular*}
\footnotesize
$q\Phi^{CV}_{B, MS}$, $q\Phi^{CV}_{B, MIS}$ = Schottky barrier heights (eV) of MS and MIS diodes respectively from C-V characteristics using general C-V method. $q\Phi^{CV,\delta}_{B, MIS}$ =  Schottky barrier heights of MIS diodes from C-V method by Cowley. $\Delta q\Phi^{CV}_B$ = $q\Phi^{CV,\delta}_{B, MIS}$ - $q\Phi^{CV}_{B, MS}$ (eV).
\end{table*}

\begin{equation}
q\Phi_B = qV_{bi} + qV_n + kT
\label{eq8}
\end{equation}
\vspace{-0.3cm} 

\begin{equation}
qV_n = E_C - E_F = {kT}ln\left(\frac{N_C}{N_D}\right)
\label{eq9}
\end{equation}
\vspace{-0.05cm}

where, $E_C$ is the conduction band minima, $E_F$ is the Fermi  level and $N_C$ is the effective density of states in the conduction band which is calculated to be 4.97$\times$10$^{18}$cm$^{-3}$ for an electron effective mass of $0.34m_o$ for $\beta$-Ga$_2$O$_3$ [10].

Although, the equation (7) in the C-V method can be a very accurate technique for Schottky barrier height extraction for metal-semiconductor SBDs, it is not appropriate for MIS SBDs \cite{Card1971, Cowley1966}. This is because it overestimates the V$_{bi}$ values for MIS structures even with very thin SiO$_2$  layers because the dielectric constant of SiO$_2$ is very low ($\epsilon_{ox}$ = 3.9$\epsilon_{o}$) and so the voltage drop across the oxide cannot be considered negligible as assumed earlier. The effect of the presence of an interfacial layer on the V$_{bi}$ extraction from (A/C)$^{2}$-V plots was  well studied in the past which shows that the oxide layer voltage drop and interface trap charges if not accounted for can lead to higher extracted values for $V_{bi}$ \cite{Card1971, Cowley1966}. Assuming the occupancy of the interface trap charges is completely governed by the semiconductor Fermi level and as the variation of interface trap state density is not too dramatic (of the same order) within the semiconductor band gap, the capacitance-voltage relationship for a reversed biased n-type MIS SBD as modeled by Cowley \cite{Cowley1966} can be expressed as,

\begin{multline}
\frac{A^2}{C^2} = \frac{2(1+\alpha)}{q\epsilon_sN_D}\left[(1+\alpha)\left(V_{bi} - \frac{kT}{q}\right) \right. \\ + \left. \sqrt{V_1\left(V_{bi} - \frac{kT}{q}\right)} + V + \frac{V_1}{4(1+\alpha)}\right]
\label{eq10}
\end{multline}
\vspace{-0.0cm} 

and, $\alpha = qD_{it}\frac{\delta}{\epsilon_{ox}}$ and $V_1 = 2q\epsilon_sN_D\frac{\delta^2}{\epsilon_{ox}^2}$
\vspace{0.5cm}

where, $\delta$    is the interfacial oxide layer thickness ($\sim$ 3nm SiO$_2$)  and  $D_{it}$   is a mean interface trap state density estimated using equation (5). The V-axis intercept voltage, $V_o$ from the linear (A/C)$^{2}$-V plots is given by,

\begin{equation}
V_o = (1+\alpha)V_{bi} + \sqrt{V_1V_{bi}} + \frac{V_1}{4(1+\alpha)}
\label{eq11}
\end{equation}

The small correction of $kT$ that arise due to mobile carriers near the depletion region edge \cite{Cowley1966} was added to the barrier height calculation like in equation (8). It can be considered that V$_{bi}$ extracted from V-axis intercept of the (A/C)$^{2}$-V plot using equation (11) to be the true V$_{bi}$ for all the MIS devices. Table II summarizes the MIS diode barrier heights extracted using both the general C-V method ($q\Phi^{CV}_{B, MIS}$) and C-V method with correction proposed by Cowley ($q\Phi^{CV,\delta}_{B, MIS}$). It can be seen that the general C-V method applied to MIS diodes overestimated the V$_{bi}$ and hence the SBH values for all the devices on three orientations by $\sim$ 0.1-0.2 V. Therefore, for MIS SBDs, only the SBH values extracted using equation (11) were considered for further analysis.

%\begin{figure}
%\centering
%  \includegraphics[width=6in,height=6cm,keepaspectratio]{table22.jpg}
%\label{fig 1}
%\end{figure}

The MS SBDs were analyzed using the general C-V method and the measured SBH ($q\Phi^{CV}_{B, MS}$) values were in the range of 0.71 eV - 1.5 eV, a bit higher than those from I-V measurements, as expected. For the MS SBDs, the highest measured barrier height (1.5 eV) was on the (010)-oriented substrate using Ni as the Schottky metal. On the (-201) and (100)-oriented substrates, the MS Pt SBDs showed higher measured barrier heights than the Ni SBDs. The (100)-oriented MS SBDs exhibited the lowest barrier heights compared to all other orientations (Pt$\colon$0.92 eV, Ni$\colon$0.71 eV). (100)-oriented $\beta$-Ga$_2$O$_3$ has consistently exhibited lower  barrier heights in literature than the other two orientations (Figure 1).

For the MIS SBDs, all the Pt MIS SBDs exhibited higher barrier heights than Ni for their respective orientations. The highest  barrier  achieved  is 1.81eV  for  the  Pt MIS SBDs on (010) substrate. Although Pt MIS SBDs on (010) and (-201)-oriented substrates exhibited considerable increment in SBH (0.53 eV and 0.37 eV respectively), but the Ni SBDs exhibited a bit lower increment in SBH (0.04 eV and 0.02 eV respectively). One possible  reason could  be that either  the FLP  effect due to  MIGS were already low on Ni SBDs or the metal electron wave functions could still be penetrating into the semiconductor bandgap through the thin SiO$_2$ interfacial layer. MIGS penetration through a high bandgap dielectric layer is highly unlikely \cite{Agrawal2012}, indicating that FLP effect was indeed lower to begin with in the case of Ni diodes. The Pt and Ni MIS SBDs exhibited large improvement in the SBHs on the (100) oriented substrates with an increment of 0.52eV and 0.61eV respectively (Pt$\colon$1.44 eV, Ni$\colon$1.32 eV). This is because of the decoupling of the Fermi level in the semiconductor and the metal due to the insertion of an interlayer dielectric. Surface-pretreatment or metal deposition in oxygen-rich conditions to reduce oxygen vacancies at the surface has been previously reported to result in some of the highest barrier heights on $\beta$-Ga$_2$O$_3$ \cite{Yao2017, Hou2019}. Apart from FLP due to MIGS penetration, oxygen vacancy defect sites at the surface of $\beta$-Ga$_2$O$_3$ has also been predicted to pin the Fermi-level  at specific energy levels (1.3 eV, 1.6 eV and 2.2 eV) below the conduction band edge \cite{Hou2019a}. Gao et. al. experimentally demonstrated that remote oxygen-plasma treatment of $\beta$-Ga$_2$O$_3$ surface can lead to diffusion of activated oxygen atoms into the lattice from the surface and thus, reduce oxygen-vacancy related defects \cite{Gao2018}. Therefore, we hypothesize that inserting a high bandgap interfacial dielectric layer (SiO$_2$) blocks MIGS penetration and remote oxygen plasma pretreatment prior to dielectric deposition could passivate oxygen vacancies at the interface which can result in enhanced Schottky barrier heights in $\beta$-Ga$_2$O$_3$.

\section{Conclusions}
\label{sec5}
In this work, we demonstrate the enhancement of Schottky barrier heights on three orientations of $\beta$-Ga$_2$O$_3$ substrates by insertion of ultra-thin SiO$_2$ interfacial layer at the MS junction. Pt and Ni MS and MIS SBDs were fabricated on three different orientations ((010), (-201) and (100)) of $\beta$-Ga$_2$O$_3$ to investigate and compare orientation dependence on barrier height modulation and these devices were characterized by room temperature I-V and C-V measurements. Pt MIS SBDs showed on average an increment of 0.37 - 0.53 eV compared to their MS counterparts. (100)-oriented $\beta$-Ga$_2$O$_3$, in general, has lower barrier heights than the other two orientations. (100)-oriented MIS SBDs showed dramatic enhancement of barrier heights (1.5$\times$ - 1.8$\times$) and reduction of reverse leakage current on this orientation due to significant enhancement of SBH with the interlayer dielectric. A promising application of this technique can be the realization of Enhancement-mode MESFETs with low gate leakage.

\section*{Acknowledgement}
This material is based upon work supported by the Air Force Office of Scientific Research under award number FA9550-18-1-0507 (Program Manager: Dr. Ali Sayir). Any opinions, finding, and conclusions or recommendations expressed in this material are those of the author(s) and do not necessarily reflect the views of the United States Air Force. This work was performed in part at the Utah Nanofab sponsored by the College of Engineering and the Office of the Vice President for Research.  We also thank Jonathan Ogle and Prof. Luisa Whittaker-Brooks at the University of Utah for providing access to equipment used in this work.

\bibliographystyle{IEEEtran}

\bibliography{main.bbl}

% Generated by IEEEtran.bst, version: 1.14 (2015/08/26)
\begin{thebibliography}{10}
\providecommand{\url}[1]{#1}
\csname url@samestyle\endcsname
\providecommand{\newblock}{\relax}
\providecommand{\bibinfo}[2]{#2}
\providecommand{\BIBentrySTDinterwordspacing}{\spaceskip=0pt\relax}
\providecommand{\BIBentryALTinterwordstretchfactor}{4}
\providecommand{\BIBentryALTinterwordspacing}{\spaceskip=\fontdimen2\font plus
\BIBentryALTinterwordstretchfactor\fontdimen3\font minus
  \fontdimen4\font\relax}
\providecommand{\BIBforeignlanguage}[2]{{%
\expandafter\ifx\csname l@#1\endcsname\relax
\typeout{** WARNING: IEEEtran.bst: No hyphenation pattern has been}%
\typeout{** loaded for the language `#1'. Using the pattern for}%
\typeout{** the default language instead.}%
\else
\language=\csname l@#1\endcsname
\fi
#2}}
\providecommand{\BIBdecl}{\relax}
\BIBdecl

\bibitem{Higashiwaki2012}
\BIBentryALTinterwordspacing
M.~Higashiwaki, K.~Sasaki, A.~Kuramata, T.~Masui, and S.~Yamakoshi, ``{Gallium
  oxide (Ga$_2$O$_3$) metal-semiconductor field-effect transistors on
  single-crystal $\beta$-Ga$_2$O$_3$ (010) substrates},'' \emph{Applied Physics
  Letters}, vol. 100, no.~1, p. 013504, 2012. [Online]. Available:
  \url{https://doi.org/10.1063/1.3674287}
\BIBentrySTDinterwordspacing

\bibitem{He2006}
\BIBentryALTinterwordspacing
H.~He, R.~Orlando, M.~A. Blanco, R.~Pandey, E.~Amzallag, I.~Baraille, and
  M.~R\'erat, ``{First-principles study of the structural, electronic, and
  optical properties of Ga$_2$O$_3$ in its monoclinic and hexagonal phases},''
  \emph{Physical Review B}, vol.~74, p. 195123, Nov 2006. [Online]. Available:
  \url{https://link.aps.org/doi/10.1103/PhysRevB.74.195123}
\BIBentrySTDinterwordspacing

\bibitem{Higashiwaki2018}
\BIBentryALTinterwordspacing
M.~Higashiwaki and G.~H. Jessen, ``Guest editorial: The dawn of gallium oxide
  microelectronics,'' \emph{Applied Physics Letters}, vol. 112, no.~6, p.
  060401, 2018. [Online]. Available: \url{https://doi.org/10.1063/1.5017845}
\BIBentrySTDinterwordspacing

\bibitem{Higashiwaki2014}
\BIBentryALTinterwordspacing
M.~Higashiwaki, K.~Sasaki, A.~Kuramata, T.~Masui, and S.~Yamakoshi,
  ``Development of gallium oxide power devices,'' \emph{physica status solidi
  (a)}, vol. 211, no.~1, pp. 21--26, 2014. [Online]. Available:
  \url{https://onlinelibrary.wiley.com/doi/abs/10.1002/pssa.201330197}
\BIBentrySTDinterwordspacing

\bibitem{Murakami2015}
\BIBentryALTinterwordspacing
H.~Murakami, K.~Nomura, K.~Goto, K.~Sasaki, K.~Kawara, Q.~T. Thieu, R.~Togashi,
  Y.~Kumagai, M.~Higashiwaki, A.~Kuramata, S.~Yamakoshi, B.~Monemar, and
  A.~Koukitu, ``{Homoepitaxial growth of $\beta$-Ga$_2$O$_3$ layers by halide
  vapor phase epitaxy},'' \emph{Applied Physics Express}, vol.~8, no.~1, p.
  015503, dec 2014. [Online]. Available:
  \url{https://doi.org/10.75672Fapex.8.015503}
\BIBentrySTDinterwordspacing

\bibitem{Sasaki2012}
\BIBentryALTinterwordspacing
K.~Sasaki, A.~Kuramata, T.~Masui, E.~G. V{\'{\i}}llora, K.~Shimamura, and
  S.~Yamakoshi, ``{Device-Quality $\beta$-Ga$_2$O$_3$ Epitaxial Films
  Fabricated by Ozone Molecular Beam Epitaxy},'' \emph{Applied Physics
  Express}, vol.~5, no.~3, p. 035502, feb 2012. [Online]. Available:
  \url{https://doi.org/10.11432Fapex.5.035502}
\BIBentrySTDinterwordspacing

\bibitem{Rafique2016}
\BIBentryALTinterwordspacing
S.~Rafique, L.~Han, M.~J. Tadjer, J.~A. Freitas, N.~A. Mahadik, and H.~Zhao,
  ``{Homoepitaxial growth of $\beta$-Ga$_2$O$_3$ thin films by low pressure
  chemical vapor deposition},'' \emph{Applied Physics Letters}, vol. 108,
  no.~18, p. 182105, 2016. [Online]. Available:
  \url{https://doi.org/10.1063/1.4948944}
\BIBentrySTDinterwordspacing

\bibitem{Shinohara2008}
\BIBentryALTinterwordspacing
Y.~Zhang, F.~Alema, A.~Mauze, O.~S. Koksaldi, R.~Miller, A.~Osinsky, and J.~S.
  Speck, ``{MOCVD grown epitaxial $\beta$-Ga$_2$O$_3$ thin film with an
  electron mobility of 176 cm$^2$/Vs at room temperature},'' \emph{APL
  Materials}, vol.~7, no.~2, p. 022506, 2019. [Online]. Available:
  \url{https://doi.org/10.1063/1.5058059}
\BIBentrySTDinterwordspacing

\bibitem{Wagner2014}
\BIBentryALTinterwordspacing
G.~Wagner, M.~Baldini, D.~Gogova, M.~Schmidbauer, R.~Schewski, M.~Albrecht,
  Z.~Galazka, D.~Klimm, and R.~Fornari, ``{Homoepitaxial growth of
  $\beta$-Ga$_2$O$_3$ layers by metal-organic vapor phase epitaxy},''
  \emph{physica status solidi (a)}, vol. 211, no.~1, pp. 27--33, 2014.
  [Online]. Available:
  \url{https://onlinelibrary.wiley.com/doi/abs/10.1002/pssa.201330092}
\BIBentrySTDinterwordspacing

\bibitem{Pearton2018}
\BIBentryALTinterwordspacing
S.~J. Pearton, J.~Yang, P.~H. Cary, F.~Ren, J.~Kim, M.~J. Tadjer, and M.~A.
  Mastro, ``{A review of Ga$_2$O$_3$ materials, processing, and devices},''
  \emph{Applied Physics Reviews}, vol.~5, no.~1, p. 011301, 2018. [Online].
  Available: \url{https://doi.org/10.1063/1.5006941}
\BIBentrySTDinterwordspacing

\bibitem{Stepanov2016}
S.~I. Stepanov, V.~I. Nikolaev, V.~E. Bougrov, and A.~E. Romanov, ``{Gallium
  Oxide : Properties and Applications - A Review},'' 2016.

\bibitem{Michaelson1977}
\BIBentryALTinterwordspacing
H.~B. Michaelson, ``The work function of the elements and its periodicity,''
  \emph{Journal of Applied Physics}, vol.~48, no.~11, pp. 4729--4733, 1977.
  [Online]. Available: \url{https://doi.org/10.1063/1.323539}
\BIBentrySTDinterwordspacing

\bibitem{Yao2017}
\BIBentryALTinterwordspacing
Y.~Yao, R.~Gangireddy, J.~Kim, K.~K. Das, R.~F. Davis, and L.~M. Porter,
  ``{Electrical behavior of $\beta$-Ga$_2$O$_3$ Schottky diodes with different
  Schottky metals},'' \emph{Journal of Vacuum Science \& Technology B},
  vol.~35, no.~3, p. 03D113, 2017. [Online]. Available:
  \url{https://doi.org/10.1116/1.4980042}
\BIBentrySTDinterwordspacing

\bibitem{Hou2019a}
\BIBentryALTinterwordspacing
C.~Hou, R.~M. Gazoni, R.~J. Reeves, and M.~W. Allen, ``{Direct comparison of
  plain and oxidized metal Schottky contacts on $\beta$-Ga$_2$O$_3$},''
  \emph{Applied Physics Letters}, vol. 114, no.~3, p. 033502, 2019. [Online].
  Available: \url{https://doi.org/10.1063/1.5079423}
\BIBentrySTDinterwordspacing

\bibitem{He2017}
\BIBentryALTinterwordspacing
Q.~He, W.~Mu, H.~Dong, S.~Long, Z.~Jia, H.~Lv, Q.~Liu, M.~Tang, X.~Tao, and
  M.~Liu, ``{Schottky barrier diode based on $\beta$-Ga$_2$O$_3$ (100) single
  crystal substrate and its temperature-dependent electrical
  characteristics},'' \emph{Applied Physics Letters}, vol. 110, no.~9, p.
  093503, 2017. [Online]. Available: \url{https://doi.org/10.1063/1.4977766}
\BIBentrySTDinterwordspacing

\bibitem{Armstrong2016}
\BIBentryALTinterwordspacing
A.~M. Armstrong, M.~H. Crawford, A.~Jayawardena, A.~Ahyi, and S.~Dhar, ``Role
  of self-trapped holes in the photoconductive gain of $\beta$-gallium oxide
  schottky diodes,'' \emph{Journal of Applied Physics}, vol. 119, no.~10, p.
  103102, 2016. [Online]. Available: \url{https://doi.org/10.1063/1.4943261}
\BIBentrySTDinterwordspacing

\bibitem{Yang2018}
J.~Yang, F.~Ren, M.~Tadjer, S.~Pearton, and A.~Kuramata, ``{2300V reverse
  breakdown voltage Ga$_2$O$_3$ schottky rectifiers},'' \emph{ECS Journal of
  Solid State Science and Technology}, vol.~7, no.~5, pp. Q92--Q96, 2018.

\bibitem{Oh2017}
S.~Oh, G.~Yang, and J.~Kim, ``{Electrical characteristics of vertical
  Ni/$\beta$-Ga$_2$O$_3$ schottky barrier diodes at high temperatures},''
  \emph{ECS Journal of Solid State Science and Technology}, vol.~6, no.~2, pp.
  Q3022--Q3025, 2017.

\bibitem{Yang2017}
\BIBentryALTinterwordspacing
J.~Yang, S.~Ahn, F.~Ren, S.~J. Pearton, S.~Jang, J.~Kim, and A.~Kuramata,
  ``{High reverse breakdown voltage Schottky rectifiers without edge
  termination on Ga$_2$O$_3$},'' \emph{Applied Physics Letters}, vol. 110,
  no.~19, p. 192101, 2017. [Online]. Available:
  \url{https://doi.org/10.1063/1.4983203}
\BIBentrySTDinterwordspacing

\bibitem{Irmscher2011}
\BIBentryALTinterwordspacing
K.~Irmscher, Z.~Galazka, M.~Pietsch, R.~Uecker, and R.~Fornari, ``{Electrical
  properties of $\beta$-Ga$_2$O$_3$ single crystals grown by the Czochralski
  method},'' \emph{Journal of Applied Physics}, vol. 110, no.~6, p. 063720,
  2011. [Online]. Available: \url{https://doi.org/10.1063/1.3642962}
\BIBentrySTDinterwordspacing

\bibitem{Suzuki2009}
\BIBentryALTinterwordspacing
R.~Suzuki, S.~Nakagomi, Y.~Kokubun, N.~Arai, and S.~Ohira, ``{Enhancement of
  responsivity in solar-blind $\beta$-Ga$_2$O$_3$ photodiodes with a Au
  Schottky contact fabricated on single crystal substrates by annealing},''
  \emph{Applied Physics Letters}, vol.~94, no.~22, p. 222102, 2009. [Online].
  Available: \url{https://doi.org/10.1063/1.3147197}
\BIBentrySTDinterwordspacing

\bibitem{Splith2014}
\BIBentryALTinterwordspacing
D.~Splith, S.~Müller, F.~Schmidt, H.~von Wenckstern, J.~J. van Rensburg, W.~E.
  Meyer, and M.~Grundmann, ``{Determination of the mean and the homogeneous
  barrier height of Cu Schottky contacts on heteroepitaxial $\beta$-Ga$_2$O$_3$
  thin films grown by pulsed laser deposition},'' \emph{physica status solidi
  (a)}, vol. 211, no.~1, pp. 40--47, 2014. [Online]. Available:
  \url{https://onlinelibrary.wiley.com/doi/abs/10.1002/pssa.201330088}
\BIBentrySTDinterwordspacing

\bibitem{Mohamed2012}
\BIBentryALTinterwordspacing
M.~Mohamed, K.~Irmscher, C.~Janowitz, Z.~Galazka, R.~Manzke, and R.~Fornari,
  ``{Schottky barrier height of Au on the transparent semiconducting oxide
  $\beta$-Ga$_2$O$_3$},'' \emph{Applied Physics Letters}, vol. 101, no.~13, p.
  132106, 2012. [Online]. Available: \url{https://doi.org/10.1063/1.4755770}
\BIBentrySTDinterwordspacing

\bibitem{Hou2019}
C.~{Hou}, R.~M. {Gazoni}, R.~J. {Reeves}, and M.~W. {Allen}, ``{Oxidized Metal
  Schottky Contacts on (010) $\beta$-Ga$_2$O$_3$},'' \emph{IEEE Electron Device
  Letters}, vol.~40, no.~2, pp. 337--340, Feb 2019.

\bibitem{Sasaki2017}
K.~{Sasaki}, D.~{Wakimoto}, Q.~T. {Thieu}, Y.~{Koishikawa}, A.~{Kuramata},
  M.~{Higashiwaki}, and S.~{Yamakoshi}, ``{First Demonstration of Ga$_2$O$_3$
  Trench MOS-Type Schottky Barrier Diodes},'' \emph{IEEE Electron Device
  Letters}, vol.~38, no.~6, pp. 783--785, June 2017.

\bibitem{Higashiwaki2016}
\BIBentryALTinterwordspacing
M.~Higashiwaki, K.~Konishi, K.~Sasaki, K.~Goto, K.~Nomura, Q.~T. Thieu,
  R.~Togashi, H.~Murakami, Y.~Kumagai, B.~Monemar, A.~Koukitu, A.~Kuramata, and
  S.~Yamakoshi, ``{Temperature-dependent capacitance-voltage and
  current-voltage characteristics of Pt/Ga$_2$O$_3$ (001) Schottky barrier
  diodes fabricated on n$^-$ - Ga$_2$O$_3$ drift layers grown by halide vapor
  phase epitaxy},'' \emph{Applied Physics Letters}, vol. 108, no.~13, p.
  133503, 2016. [Online]. Available: \url{https://doi.org/10.1063/1.4945267}
\BIBentrySTDinterwordspacing

\bibitem{Jian2018}
\BIBentryALTinterwordspacing
G.~Jian, Q.~He, W.~Mu, B.~Fu, H.~Dong, Y.~Qin, Y.~Zhang, H.~Xue, S.~Long,
  Z.~Jia, H.~Lv, Q.~Liu, X.~Tao, and M.~Liu, ``{Characterization of the
  inhomogeneous barrier distribution in a Pt/(100)$\beta$-Ga$_2$O$_3$ Schottky
  diode via its temperature-dependent electrical properties},'' \emph{AIP
  Advances}, vol.~8, no.~1, p. 015316, 2018. [Online]. Available:
  \url{https://doi.org/10.1063/1.5007197}
\BIBentrySTDinterwordspacing

\bibitem{Farzana2017}
\BIBentryALTinterwordspacing
E.~Farzana, Z.~Zhang, P.~K. Paul, A.~R. Arehart, and S.~A. Ringel, ``{Influence
  of metal choice on (010) $\beta$-Ga$_2$O$_3$ Schottky diode properties},''
  \emph{Applied Physics Letters}, vol. 110, no.~20, p. 202102, 2017. [Online].
  Available: \url{https://doi.org/10.1063/1.4983610}
\BIBentrySTDinterwordspacing

\bibitem{Konishi2017}
\BIBentryALTinterwordspacing
K.~Konishi, K.~Goto, H.~Murakami, Y.~Kumagai, A.~Kuramata, S.~Yamakoshi, and
  M.~Higashiwaki, ``{1-kV vertical Ga$_2$O$_3$ field-plated Schottky barrier
  diodes},'' \emph{Applied Physics Letters}, vol. 110, no.~10, p. 103506, 2017.
  [Online]. Available: \url{https://doi.org/10.1063/1.4977857}
\BIBentrySTDinterwordspacing

\bibitem{Jiang2019}
\BIBentryALTinterwordspacing
K.~Jiang, L.~A. Lyle, E.~Favela, D.~Moody, T.~Lin, K.~K. Das, A.~Popp,
  Z.~Galazka, G.~Wagner, and L.~M. Porter, ``{Electrical Properties of (100)
  $\beta$-Ga$_2$O$_3$ Schottky Diodes with Four Different Metals},'' \emph{ECS
  Transactions}, vol.~92, no.~7, pp. 71--78, 2019. [Online]. Available:
  \url{https://doi: 10.1149/09207.0071ecst}
\BIBentrySTDinterwordspacing

\bibitem{Jang2017}
\BIBentryALTinterwordspacing
S.~Jang, S.~Jung, K.~Beers, J.~Yang, F.~Ren, A.~Kuramata, S.~Pearton, and K.~H.
  Baik, ``{A comparative study of wet etching and contacts on ($\bar{\sf2}01$)
  and (010) oriented $\beta$-Ga$_2$O$_3$},'' \emph{Journal of Alloys and
  Compounds}, vol. 731, pp. 118 -- 125, 2018. [Online]. Available:
  \url{http://www.sciencedirect.com/science/article/pii/S0925838817333984}
\BIBentrySTDinterwordspacing

\bibitem{Fu2018}
H.~{Fu}, H.~{Chen}, X.~{Huang}, I.~{Baranowski}, J.~{Montes}, T.~{Yang}, and
  Y.~{Zhao}, ``{A Comparative Study on the Electrical Properties of Vertical
  ($\bar{\sf2}01$) and (010) $\beta$-Ga$_2$O$_3$ Schottky Barrier Diodes on EFG
  Single-Crystal Substrates},'' \emph{IEEE Transactions on Electron Devices},
  vol.~65, no.~8, pp. 3507--3513, Aug 2018.

\bibitem{Robertson2006}
\BIBentryALTinterwordspacing
J.~Robertson and B.~Falabretti, ``{Band offsets of high K gate oxides on III-V
  semiconductors},'' \emph{Journal of Applied Physics}, vol. 100, no.~1, p.
  014111, 2006. [Online]. Available: \url{https://doi.org/10.1063/1.2213170}
\BIBentrySTDinterwordspacing

\bibitem{Monch1999}
\BIBentryALTinterwordspacing
W.~Monch, ``Barrier heights of real schottky contacts explained by
  metal-induced gap states and lateral inhomogeneities,'' \emph{Journal of
  Vacuum Science \& Technology B: Microelectronics and Nanometer Structures
  Processing, Measurement, and Phenomena}, vol.~17, no.~4, pp. 1867--1876,
  1999. [Online]. Available:
  \url{https://avs.scitation.org/doi/abs/10.1116/1.590839}
\BIBentrySTDinterwordspacing

\bibitem{Roy2010}
A.~M. {Roy}, J.~Y.~J. {Lin}, and K.~C. {Saraswat}, ``Specific contact
  resistivity of tunnel barrier contacts used for fermi level depinning,''
  \emph{IEEE Electron Device Letters}, vol.~31, no.~10, pp. 1077--1079, Oct
  2010.

\bibitem{Lin2011}
\BIBentryALTinterwordspacing
J.-Y.~J. Lin, A.~M. Roy, A.~Nainani, Y.~Sun, and K.~C. Saraswat, ``{Increase in
  current density for metal contacts to n-germanium by inserting TiO$_2$
  interfacial layer to reduce Schottky barrier height},'' \emph{Applied Physics
  Letters}, vol.~98, no.~9, p. 092113, 2011. [Online]. Available:
  \url{https://doi.org/10.1063/1.3562305}
\BIBentrySTDinterwordspacing

\bibitem{Agrawal2012}
\BIBentryALTinterwordspacing
A.~Agrawal, N.~Shukla, K.~Ahmed, and S.~Datta, ``{A unified model for insulator
  selection to form ultra-low resistivity metal-insulator-semiconductor
  contacts to n-Si, n-Ge, and n-InGaAs},'' \emph{Applied Physics Letters}, vol.
  101, no.~4, p. 042108, 2012. [Online]. Available:
  \url{https://doi.org/10.1063/1.4739784}
\BIBentrySTDinterwordspacing

\bibitem{Agrawal2014}
\BIBentryALTinterwordspacing
A.~Agrawal, J.~Lin, M.~Barth, R.~White, B.~Zheng, S.~Chopra, S.~Gupta, K.~Wang,
  J.~Gelatos, S.~E. Mohney, and S.~Datta, ``Fermi level depinning and contact
  resistivity reduction using a reduced titania interlayer in n-silicon
  metal-insulator-semiconductor ohmic contacts,'' \emph{Applied Physics
  Letters}, vol. 104, no.~11, p. 112101, 2014. [Online]. Available:
  \url{https://doi.org/10.1063/1.4868302}
\BIBentrySTDinterwordspacing

\bibitem{Saleh2019}
\BIBentryALTinterwordspacing
M.~Saleh, A.~Bhattacharyya, J.~B. Varley, S.~Swain, J.~Jesenovec,
  S.~Krishnamoorthy, and K.~Lynn, ``{Electrical and optical properties of Zr
  doped $\beta$-Ga$_2$O$_3$ single crystals},'' \emph{Applied Physics Express},
  vol.~12, no.~8, p. 085502, jul 2019. [Online]. Available:
  \url{https://doi.org/10.75672F1882-07862Fab2b6c}
\BIBentrySTDinterwordspacing

\bibitem{Jia2015}
\BIBentryALTinterwordspacing
Y.~Jia, K.~Zeng, J.~S. Wallace, J.~A. Gardella, and U.~Singisetti,
  ``{Spectroscopic and electrical calculation of band alignment between atomic
  layer deposited SiO$_2$ and $\beta$-Ga$_2$O$_3$ (-201)},'' \emph{Applied
  Physics Letters}, vol. 106, no.~10, p. 102107, 2015. [Online]. Available:
  \url{https://doi.org/10.1063/1.4915262}
\BIBentrySTDinterwordspacing

\bibitem{Sze2006}
S.~Sze and K.~K. Ng, \emph{{Physics of Semiconductor Devices}}.\hskip 1em plus
  0.5em minus 0.4em\relax John Wiley {\&} Sons, Inc., 2006.

\bibitem{He2018}
Q.~{He}, W.~{Mu}, B.~{Fu}, Z.~{Jia}, S.~{Long}, Z.~{Yu}, Z.~{Yao}, W.~{Wang},
  H.~{Dong}, Y.~{Qin}, G.~{Jian}, Y.~{Zhang}, H.~{Xue}, H.~{Lv}, Q.~{Liu},
  M.~{Tang}, X.~{Tao}, and M.~{Liu}, ``{Schottky Barrier Rectifier Based on
  (100) $\beta$-Ga$_2$O$_3$ and its DC and AC Characteristics},'' \emph{IEEE
  Electron Device Letters}, vol.~39, no.~4, pp. 556--559, April 2018.

\bibitem{Werner1991}
\BIBentryALTinterwordspacing
J.~H. Werner and H.~H. Güttler, ``{Barrier inhomogeneities at Schottky
  contacts},'' \emph{Journal of Applied Physics}, vol.~69, no.~3, pp.
  1522--1533, 1991. [Online]. Available: \url{https://doi.org/10.1063/1.347243}
\BIBentrySTDinterwordspacing

\bibitem{Card1971}
\BIBentryALTinterwordspacing
H.~C. Card and E.~H. Rhoderick, ``{Studies of tunnel {MOS} diodes I. Interface
  effects in silicon Schottky diodes},'' \emph{Journal of Physics D: Applied
  Physics}, vol.~4, no.~10, pp. 1589--1601, oct 1971. [Online]. Available:
  \url{https://doi.org/10.10882F0022-37272F42F102F319}
\BIBentrySTDinterwordspacing

\bibitem{Cowley1966}
\BIBentryALTinterwordspacing
A.~M. Cowley, ``{Depletion Capacitance and Diffusion Potential of Gallium
  Phosphide Schottky-Barrier Diodes},'' \emph{Journal of Applied Physics},
  vol.~37, no.~8, pp. 3024--3032, 1966. [Online]. Available:
  \url{https://doi.org/10.1063/1.1703157}
\BIBentrySTDinterwordspacing

\bibitem{Zeng2017}
\BIBentryALTinterwordspacing
K.~Zeng and U.~Singisetti, ``{Temperature dependent quasi-static
  capacitance-voltage characterization of SiO$_2$/$\beta$-Ga$_2$O$_3$ interface
  on different crystal orientations},'' \emph{Applied Physics Letters}, vol.
  111, no.~12, p. 122108, 2017. [Online]. Available:
  \url{https://doi.org/10.1063/1.4991400}
\BIBentrySTDinterwordspacing

\bibitem{Gao2018}
\BIBentryALTinterwordspacing
H.~Gao, S.~Muralidharan, N.~Pronin, M.~R. Karim, S.~M. White, T.~Asel,
  G.~Foster, S.~Krishnamoorthy, S.~Rajan, L.~R. Cao, M.~Higashiwaki, H.~von
  Wenckstern, M.~Grundmann, H.~Zhao, D.~C. Look, and L.~J. Brillson, ``{Optical
  signatures of deep level defects in Ga$_2$O$_3$},'' \emph{Applied Physics
  Letters}, vol. 112, no.~24, p. 242102, 2018. [Online]. Available:
  \url{https://doi.org/10.1063/1.5026770}
\BIBentrySTDinterwordspacing

\end{thebibliography}

\end{document}